\documentclass[twocolumn]{aastex631}

\usepackage{amsmath,amssymb}
\usepackage{graphicx}
\usepackage{bm}
\usepackage{xcolor}
\begin{document}
\title{Redshift Identifiability from Gamma‑Ray Burst Prompt Emission}

\author{Shu-Xu Yi}
\affiliation{State Key Laboratory of Particle Astrophysics, Institute of High Energy Physics, Chinese Academy of Sciences, Beijing 100049, People's Republic of China}
\affiliation{University of Chinese Academy of Sciences, Beijing 100049, China}
\email{sxyi@ihep.ac.cn}

\begin{abstract}
We examine which prompt-emission observables of gamma-ray bursts (GRBs) carry identifiable redshift information in realistic structured-jet scenarios. By examining the photon-level Lorentz and cosmological transformations, we demonstrate any spectral observable reflecting the dominant-patch emission and any temporal observable reflecting the co-moving frame physics in the jet suffers from the same $z$-$\theta_v$ degeneracy. The observable reflecting the absolute flux or count rate will also suffer from this $z$-$\theta_v$ degeneracy in a different form. Due to the large variation of the intrinsic luminosity, combining the observables in the above-mentioned three classes cannot break this degeneracy. However, we demonstrate that two other classes of observables contain the potential to break this degeneracy: (i)~central engine-imprinted time scales, (ii)~emission from out of the dominant patch.

\end{abstract}

\section{Introduction}

Gamma-ray burst (GRB) redshifts are essential for characterizing burst energetics,
the cosmic star-formation history, and the evolution of the GRB population \citep{WandermanPiran2010,Perley2016,Salafia2023,2025MNRAS.541..798D}.
Yet secure spectroscopic redshifts require expensive afterglow follow-up and
are available for only a minor fraction of detected bursts. In {\it Swift}/BAT samples, only about $28\%$ of short/type-I GRBs and about $31\%$ of long/type-II GRBs have measured redshifts \citep{Horvath2016}, while in the {\it Fermi}/GBM spectral catalog the fraction with known redshift is only about $6\%$ \citep{Poolakkil2021}. The redshift incompleteness causes strong selection bias in GRB population studies, including the cosmic evolution of their progenitors and the GRB luminosity function \citep{WandermanPiran2010,Perley2016}. 
This has motivated persistent efforts to extract redshift estimates from the
prompt $\gamma$-ray emission alone---so-called pseudo-redshift methods---exploiting
empirical correlations \citep{Amati2002,Atteia2003,Yonetoku2004,Dainotti2020}. However, the feasibility of prompt redshift inference remains debated \citep{2025ApJ...981..197Y}, and degeneracy between redshift and viewing angle in structured jet in some prompt emission observables has been well-noted (see for instance, \citealt{2007MNRAS.374L..20L}). In recent years, attempts have been made to employ machine-learning based inferences \citep{2025A&A...698A..92N,2026Univ...12...31B}, to recover redshift information without explicitly using specific observables. 


As we will show systematically in this letter, most of the information from the prompt emission encodes the redshift in a way that is degenerate with the viewing angle and irrecoverable. Only some certain classes of observables have the potential to recover the redshift. 

The fundamental information from the prompt emission is the count of photons in energy-time parameter space in the observer's frame. Therefore, in the next section, we study how the redshift enters the observed energy, time-of-arrival and the count of photons. We further show the limit of identifiability of redshift from three classes of observables, together with their combination, deduced from the above-mentioned three fundamental information; In section 3, we propose two classes of observables which have the ability to break the above limitation, and testable predictions to distinguish them from the other classes; We discuss and summarize our conclusion in sections 4 and 5.

\section{How does redshift imprint in the GRB prompt emission?}
The root of prompt emission observable is the differential distribution of photons: $Nf(E_{\rm obs}, t_{\rm obs})$, where $N$ is the total number of photons, and $f$ is the normalized number density distribution, or equivalently, $E_{\rm tot} f(E_{\rm obs}, t_{\rm obs})$, where $E_{\rm tot}$ is the total deposited energy of the photons in the detector, and $f$, in this case, is the energy density distribution. $t_{\rm obs}$ is the time of arrival of photons, and $t_{\rm obs}=\int dt_{\rm obs}$. Any quantity derived from the prompt emission data depends on $E_{\rm tot}$ and $f$: $x(E_{\rm tot}, f)$. The $z$ dependency of $x$ can be written as:
\begin{widetext}
\[
\frac{dx}{dz} = \frac{\partial x}{\partial E_{\rm tot}} \frac{dE_{\rm tot}}{dz}
+ \frac{1}{N} \sum_{i=1}^N \left[
\frac{\partial g}{\partial E_{\rm obs}}\bigg|_{(E_{{\rm obs},i}, t_{{\rm obs},i})} \frac{dE_{{\rm obs},i}}{dz}
+ \frac{\partial g}{\partial t_{\rm obs}}\bigg|_{(E_{{\rm obs},i}, t_{{\rm obs},i})} \frac{dt_{{\rm obs},i}}{dz}
\right],
\label{eq:dz}
\]
\end{widetext}
where $g(E_{\rm obs}, t_{\rm obs})\equiv \frac{\delta x}{\delta f(E_{\rm obs}, t_{\rm obs})}$ is the functional derivative of $x$ with respect to the distribution $f$. The above equation formalizes an intuitive conclusion: the identifiability of redshift from the prompt emission data relies on the redshift imprint in the observed energy ($E_{\rm obs}$), time-of-arrival of each photon ($t_{\rm obs}$), and their total energy ($E_{\rm tot}$).

In the observer frame, a photon emitted with comoving energy $E'$ and comoving
time interval $dt'$ is received as
\begin{equation}
 E_{\rm obs} = \frac{\mathcal{D}}{1+z}\,E', \qquad
 dt_{\rm obs} = \frac{1+z}{\mathcal{D}}\,dt',
 \label{eq:basic}
\end{equation}
where $\mathcal{D}$ is the Doppler factor of emitting material and $z$ is the cosmological redshift of the source. The $E^\prime$ should be determined by the local dissipation and radiation physics, and independent of $z$ and $\theta_v$. For $dt_{\rm obs}$, two distinct physical origins are possible. It can be determined by the local dissipation and radiation physics \citep{2009MNRAS.394L.117N} (we denote $dt^{\rm jet}_{\rm obs}$), or by the central engine activity \citep{1997ApJ...490...92K} ($dt^{\rm eng}_{\rm obs}$). In the former case, $dt^\prime$ (the comoving interval) is independent of $z$ and $\theta_v$. In the latter case, $dt^\prime=\mathcal{D}dt_{\rm eng}$ and therefore
\begin{equation}
 dt^{\rm eng}_{\rm obs}=(1+z)dt_{\rm eng}. \label{eq:dt_eng}
\end{equation}

The bolometric fluence is determined by:
\begin{equation}
 E_{\rm tot} \propto \frac{\left<\epsilon\,\mathcal{D}^2\right>}{D_L(z)^2},
 \label{eq:Etot}
\end{equation}
where $<\cdots>$ denotes the contribution from all emitting patches, and $\epsilon$ is the jet co-moving frame emissivity.

A dominant contribution to the observed flux arises from the patch of the jet which moves along the line of sight (LOS), and therefore $\mathcal{D}\sim2\Gamma_v$, $\Gamma_v$ is the bulk Lorentz factor of the jet at the viewing angle $\theta_v$. In the dominant patch approximation (DPA),

\begin{eqnarray}
 E_{\rm obs} &\propto& \displaystyle\frac{\Gamma_v}{1+z}, \label{eq:dpa_E}\\
 t^{\rm jet}_{\rm{obs}} &\propto& \displaystyle\frac{1+z}{\Gamma_v}, \label{eq:dpa_t}\\
 E_{\rm tot} &\propto& \displaystyle\frac{\epsilon_v\,\Gamma_v^2}{D_L(z)^2}. \label{eq:dpa_F}
\end{eqnarray}

From the equations above, we see that, under DPA, for any observable constructed from the spectral shape (Class~I) and the temporal structure related to local jet physics (Class~II), as well as from their combination, the redshift \(z\) is imprinted through the combined quantity $x_E(\theta_v,z)\equiv\Gamma_v/(1+z)$. In addition, for observable related to the total energy output (Class~III), the redshift dependence appears through $x_F(\theta_v,z)\equiv\epsilon_v \Gamma_v^2 / D_L(z)^2$. We will discuss later which commonly used observable fall into each of categories.

To quantitatively demonstrate the $z$--$\theta_v$ degeneracy, we map the allowed regions in the $(z, \theta_v)$ parameter space that are consistent with representative observed values. We consider three widely used jet structure models:

\begin{itemize}
 \item \textbf{Top-hat jet:} $\Gamma(\theta) = \Gamma_0$ and $\epsilon(\theta) = \epsilon_0$ for $\theta \leq \theta_c$; zero otherwise.
 \item \textbf{Gaussian jet:} $\Gamma(\theta) = 1 + (\Gamma_0-1)\exp[-\theta^2/(2\theta_c^2)]$, $\epsilon(\theta) = \epsilon_0 \exp[-\theta^2/(2\theta_c^2)]$.
 \item \textbf{Power-law jet:} $\Gamma(\theta) = 1 + (\Gamma_0-1)/[1+(\theta/\theta_c)^s]$, $\epsilon(\theta) = \epsilon_0/[1+(\theta/\theta_c)^s]$.
\end{itemize}

Assuming the true ($z$, $\theta_v$), we compute the allowed region in the $(z, \theta_v)$ plane that can reproduce the same observed values in three classes of observables. To isolate the intrinsic $z$--$\theta_v$ degeneracy, we assume that the on-axis Lorentz factor $\Gamma_0$ is independently constrained with high precision, and adopt a fiducial prior of $\Gamma_0 = 200 \pm 50$. The $\epsilon_0$ is allowed to vary two orders of magnitude, consistent with the observed luminosity range.

\begin{figure*}[t]
 \includegraphics[width=\textwidth]{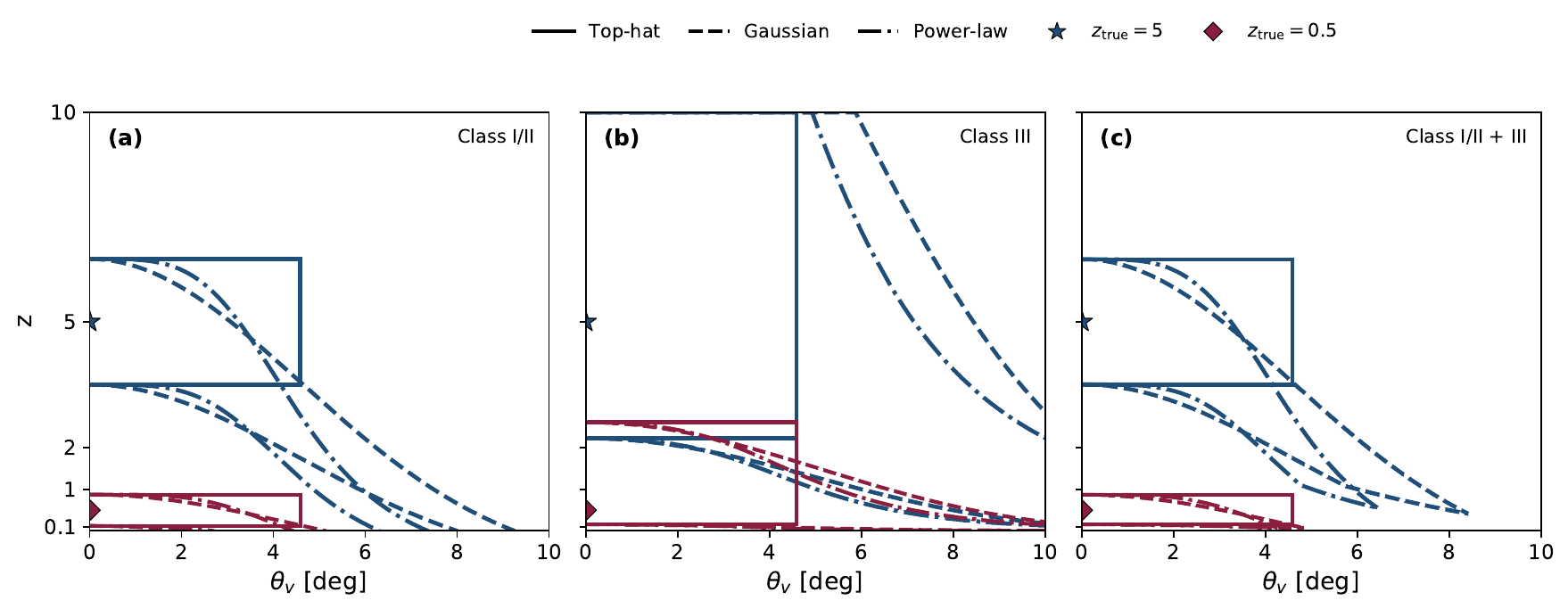}
\caption{Allowed regions in the $(z,\,\theta_v)$ plane for two fiducial sources, shown in dark blue for $(z,\theta_v)=(5,0^\circ)$ and in dark red for $(z,\theta_v)=(0.5,0^\circ)$, under three jet structure models (solid: top-hat; dashed: Gaussian; dash-dotted: power-law). Panel~(a) shows the constraint from Class~I/II observables, which are derived from spectral energy distribution or jet-physics related time scales under DPA. Panel~(b) shows the constraint from Class~III observables, which are derived from total number or deposited energy of photons. Panel~(c) shows the joint constraint from Class~I\&III observables.}
 \label{fig:main}
\end{figure*}

In Fig. \ref{fig:main}, we show the allowed regions for using class I \& II observables, class III observables, and their combination. In producing these plots, we adopt the following parameters for the jet structure models: $\theta_c=0.08$~rad for all models, $s=4$ for the power-law jet. We show two fiducial sources: a high-$z$ source at $(z, \theta_v) = (5, 0^\circ)$ and a low-$z$ source at $(z, \theta_v) = (0.5, 0^\circ)$. The allowed regions in figure \ref{fig:main} are calculated as follows: for panel a, the upper/lower boundary of the allowed region corresponds to $(\theta_v,z)$ which satisfies $x_F(\theta_v,z,\Gamma_{\rm{max/min}})=x_{F,\rm{true}}$ respectively; for panel b, the whole $(\theta_v,z)$ grids are scanned to check whether a pair of $(\Gamma_0,\epsilon_0)$ can be found within the region $[\Gamma_{0,\rm{min}},\Gamma_{0,\rm{max}}]\times[\epsilon_{0,\rm{min}},\epsilon_{0,\rm{max}}]$, such that $x_F(\theta_{v,i},z_{v,i},\Gamma_0,\epsilon_0)=x_{F,\rm{true}}$. A point $(\theta_{v,i},z_i)$ is marked allowed only when such a pair of $(\Gamma_0,\epsilon_0)$ exists; panel c shows the overlap of the allowed region from panels a and b.

From Fig.~\ref{fig:main} we can see that the $z$--$\theta_v$ degeneracy manifests clearly for the three classes of observables. In panel~(a), using only a Class~I or/and II observable, the top-hat jet shows no dependence on $\theta_v$ within the core, leading to a narrow horizontal allowed band primarily set by the variation of $\Gamma_0$. By contrast, for the Gaussian and power-law structured jets, the smooth variation of $\Gamma_v(\theta_v)$ produces an extended degeneracy in the $(z,\theta_v)$ plane, such that a high-$z$ on-axis burst and a low-$z$ off-axis burst can yield the same Class~I/II observable. In panel~(b), using only a Class~III observable, the constraint on $z$ remains weak even for the top-hat jet because of the broad on-axis luminosity range. For structured jets, the additional dependence on $\epsilon_v(\theta_v)$ and $\Gamma_v^2$ further enlarges the allowed region, especially at large viewing angles. In panel~(c), combining Class~I/II and Class~III observables reduces the allowed region to some extent, since their dependencies on $z$ and $\theta_v$ are different. Nevertheless, a substantial degeneracy remains. For the high-$z$ source, the allowed region extends from its true value to zero. While for the source with low true $z$ and small true $\theta_v$, the degeneracy is still present, but its impact on the inferred $z$ range remains limited because of the boundary of the allowed parameter space. Overall, Fig.~\ref{fig:main} shows that, for realistic jet structures, the redshift identifiability of Class~I, Class~II, and Class~III observables, as well as of their combinations, is intrinsically limited by degeneracy with the viewing angle. Except for sources that are truly at low redshift and viewed nearly on-axis, the redshift information cannot be robustly recovered from prompt observables alone. In reality, $\Gamma_0$ is poorly constrained for most GRBs. This further strengthens our conclusion on the redshift non-identifiability from Class~I, II, and III.

On the other hand, if an observed timescale reflects an intrinsic central-engine timescale, such as the duration of, or interval between, episodes of central-engine activity, then $dt_{\rm obs}^{\rm eng}$ should scale directly with $(1+z)$, without any dependence on $\Gamma_v$ or $\theta_v$. We denote observables associated with $dt_{\rm obs}^{\rm eng}$ as Class~IV.

\section{How high-latitude emission helps to break the $\theta_v$--$z$ degeneracy}
\label{sec:HLE}

The above analysis is based on DPA. Here we show that if contributions from off-LOS patches are included, the Doppler factor $\mathcal{D}$ and the emissivity $\epsilon$ evolve with observer time in a way that carries explicit $\theta_v$ dependence, enabling high-latitude emission (HLE) signatures to break the $\theta_v$--$z$ degeneracy.

We define three angles to describe the geometry:
\begin{itemize}
\item $\alpha$ -- angle between the emitting patch and the LOS (latitude with respect to the LOS);
\item $\varphi$ -- azimuthal angle around the LOS, with $\varphi=0$ aligned with the plane containing the jet axis and the LOS;
\item $\theta$ -- polar angle of the emitting patch measured from the jet axis.
\end{itemize}

By the spherical law of cosines:
\begin{equation} \cos\theta = \cos\theta_v\,\cos\alpha + \sin\theta_v\,\sin\alpha\,\cos\varphi . \label{eq:cos_theta}
\end{equation} For a photon emitted at $(\alpha,\varphi)$, the Doppler factor is \begin{equation} \mathcal{D}(\alpha,\varphi) = \frac{1} {\Gamma\bigl[\theta(\alpha,\varphi,\theta_v)\bigr]\, \bigl(1-\beta\cos\alpha\bigr)} . \end{equation}
A photon emitted at latitude $\alpha$ arrives at the observer with a time delay $\delta t_{\rm obs}$ relative to the photon from the LOS patch simultaneously:
\begin{equation} \delta t_{\rm obs}(\alpha) = (1+z)\frac{r_0}{c}\,(1-\cos\alpha) . \label{eq:dt_obs}
\end{equation}
Inverting Eq.~\ref{eq:dt_obs}, the Equal-Arrival-Time Surface (EATS) at delay $\delta t_{\rm obs}$ corresponds to the latitude
\begin{equation} \alpha_t = \arccos\!\Bigl[1-\frac{c\,\delta t_{\rm obs}}{r_0(1+z)}\Bigr] . \end{equation}
and similarly the emissivity $\epsilon(\alpha_t,\varphi)\equiv\epsilon[\theta(\alpha_t,\varphi,\theta_v)]$ is time-dependent through $\alpha_t$.

The observed specific flux at time $t_{\rm obs}$ and observed frequency $\nu_{\rm obs}$ is the sum over all patches on the EATS at $\alpha_t$:

\begin{widetext}
\[F_{\rm obs}(t_{\rm obs},\nu_{\rm obs}|\theta_v) = \int_{0}^{2\pi} \frac{\mathcal{D}^3(\alpha_t,\varphi,\theta_v)}{1+z}\, \epsilon\!\Bigl(\alpha_t,\varphi,\, \nu'=\frac{\nu_{\rm obs}}{\mathcal{D}(\alpha_t,\varphi,\theta_v)}\Bigr)\, \frac{r_0^2}{D_L^2}\, d\varphi , \label{eq:F_obs} 
\]
\end{widetext}

where $r_0$ is the emission radius.

Consequently, the time evolution of $F_{\rm obs}(\nu_{\rm obs})$ carries an explicit imprint of $\theta_v$. This is why HLE signatures have the potential to discriminate different viewing angles \citep{Ascenzi2020}. For example, \cite{2016MNRAS.461.3607S} predicted viewing-angle-dependent pulse broadening, spectral softening, and modified hardness-intensity correlation. These HLE signatures have been demonstrated in simulations \citep{2016MNRAS.461.3607S} and provide a viable path to breaking the $z$--$\theta_v$ degeneracy when combined with simulation-based training. Therefore, the \(\theta_v\)--\(z\) degeneracy can, in principle, be broken by signatures of HLE, such as the morphology and/or spectral evolution of individual pulses \citep{2016MNRAS.461.3607S}. We denote such observables as Class~V.

\section{Discussion}

Commonly used prompt observables in GRB studies can be organized into these classes according to the physical information they primarily encode. Class~I mainly includes time-integrated spectral quantities that scale with Doppler or redshift, such as the observed peak energy $E_{\rm p}$, spectral break energies, and hardness ratios. Class~II includes observables that characterize the temporal structure associated with local jet physics, such as pulse widths, rise times, and in many cases, spectral lags and minimum variability timescales when these are interpreted as reflecting the intrinsic timescale of emission from the dominant patch. Class~III includes observables related to the total energy output, such as the observed fluence and peak flux. Class~IV includes observables that are linked to intrinsic central-engine timescales, such as the durations of individual engine activity episodes, the intervals between ejecta launch and the intrinsic periodicity of the central engine \citep{2025NatAs...9.1701C}. Class~V includes observables that probe off-LOS emission patches, such as the spectral evolution in late-time prompt emission, the temporal and spectral behavior of individual pulse tails \cite{2016MNRAS.461.3607S,2026ApJ..1004...75Y}, temporal evolution of spectral lines \cite{2024SCPMA..6789511Z,2024ApJ...973L..17Z}.

Under different prompt-emission models, the same observationally defined timescale may belong to different classes. For example, the interval or waiting time between adjacent pulses can be classified as Class~IV in the internal-shock framework, where it directly reflects the time separation between shells launched by the central engine \citep{Daigne1998}. In contrast, in magnetic reconnection scenarios, the same quantity may instead be governed by local jet properties, such as the level of MHD turbulence or the degree of magnetization \citep{ZhangYan2011}, and should therefore be interpreted as a Class~II observable. This ambiguity also suggests a possible observational route for diagnosing the underlying prompt-emission mechanism. Specifically, for a GRB population with independently and reliably measured redshifts, one may examine the correlation between such a characteristic timescale, $\delta t$, and redshift. If a subset of bursts follows the expected $(1+z)$ scaling, this would indicate that, for those events, $\delta t$ most likely traces an intrinsic central-engine timescale, consistent with a Class~IV interpretation and with prompt-emission models analogous to internal shocks. By contrast, if another subset does not follow this redshift scaling, then $\delta t$ is more likely to belong to Class~II. In that case, one may further test whether $\delta t$ correlates strongly with representative Class~I observables, such as $E_{\rm p}$ or $E_{\rm cut}$. A strong correlation of this kind would support the idea that the relevant timescale is controlled by local dissipation physics, as expected in prompt-emission scenarios akin to magnetic reconnection.

The analysis above addresses the \emph{intrinsic} identifiability problem:
even with perfect, bias-free measurements, the $z$--$\theta_v$ degeneracy limits
what can be recovered. Real data introduce an additional layer of complication: instrumental selection
effects, notably the ``tip-of-the-iceberg'' effect
\citep{2022ApJ...927..157M,2026ApJ..1002...27M}.
As the distance to a GRB increases, a progressively larger fraction of its
prompt emission falls below the instrumental background.
This selection bias distorts the observed spectral shape (only the hardest
peak-flux photons are detected), the inferred isotropic energy, and the
measured duration in ways that deviate significantly from the simple $(1+z)$
cosmological scalings discussed here.

Our conclusion states that Class~I--III observables from the prompt emission cannot recover redshift. A common practice in the community is to use the Amati relation ($E_{\rm p}$--$E_{\rm iso}$) or the Yonetoku relation ($E_{\rm p}$--$L_{\rm iso}$) to estimate pseudo-redshifts. A critique of this practice is a natural extension of our conclusion. Such criticism has existed for some time: for instance, \cite{2007MNRAS.374L..20L} argued that the Amati relation is not genuine but purely a selection bias---in a flux-limited sample, more distant GRBs tend to be found on-axis, yielding higher $E_{\rm iso}$ and $E_{\rm p}$ in a structured jet, while closer GRBs tend to be found at larger viewing angles due to the larger solid angle, resulting in lower $E_{\rm iso}$ and lower $E_{\rm p}$. Furthermore, \citep{2025ApJ...981..197Y} showed that whether an empirical relation can yield a non-degenerate solution for pseudo-redshift depends sensitively on the functional form of that relation. The empirical relations applied to prompt emission, however, may have been using an over-simplified analytical form, or the wrong shape, when extrapolated from a subsample of GRBs with known redshifts to the full population.

\section{Summary}

In this letter, we examined the identifiability of GRB redshift from prompt-emission observables within a structured-jet framework. Under the dominant-patch approximation, we showed that the redshift dependence of prompt observables is generally entangled with jet geometry through the viewing-angle-dependent Doppler factor. In particular, Class~I\&II observables, and their combinations encode redshift through the combination $\Gamma_v/(1+z)$, while Class~III observables depend on $\epsilon_v \Gamma_v^2 / D_L^2$. As a result, for realistic jet structures, prompt observables alone generally cannot provide a determination of redshift. This intrinsic $z$--$\theta_v$ degeneracy persists even when multiple classes of observables are combined, except in the limited case of bursts that are intrinsically nearby and viewed close to the jet axis.

On the other hand, if an observed timescale directly traces the intrinsic cadence of central-engine activity, it belongs to Class~IV and scales with $(1+z)$ without degeneracy with viewing-angle. Furthermore, observables sensitive to high-latitude emission and off-LOS contributions belong to Class~V, and may in principle help break the $z$--$\theta_v$ degeneracy by probing jet structure.

We also discussed that the class assignment of some commonly used timing observables may depend on the underlying prompt-emission mechanism. For example, pulse intervals or waiting times may trace engine activity in internal-shock-like scenarios, but may instead reflect local jet dissipation physics in magnetic reconnection scenarios. This suggests that population studies of prompt timescales, when combined with independent redshift measurements, may provide a useful way to diagnose the physical origin of prompt variability.

Particularly, we suggest tests on the archival data: Construct large GRB samples with independently measured redshifts and prompt timing observables, \textit{e.g.,} pulse intervals, pulses waiting-time; Test whether different prompt timescales follow the expected $(1+z)$ scaling across the population, in order to identify which observables behave as Class~IV tracers of intrinsic central-engine activity. Previous studies of (1+z) scaling in GRB timing observables have yielded inconsistent results: \cite{2013ApJ...765..116K} found no time dilation signatures in Swift/BAT data, while \cite{Littlejohns2014} reported an increasing-duration trend with redshift in Fermi/GBM data, though selection effects complicate the interpretation. This tension is expected: the presence or absence of an apparent (1+z) scaling in any small sample would depend on the ratio of bursts in which such timescales are Classes II or IV. Search for correlations between candidate Class~II timescales and representative Class~I observables, such as $E_{\rm p}$ or $E_{\rm cut}$, to probe whether local dissipation physics controls the prompt variability.

For Class V, the path forward is simulation-based machine learning guided by this analysis. Our work shows that the $\theta_v$ information resides specifically in the high-latitude-emission regions of the light curve. Unlike blind ML training on the full light curve -- which may miss or misweight the $\theta_v$-sensitive features -- a physics-informed approach first trains on the HLE regions to extract $\theta_v$, then uses it to break the $z$--$\theta_v$ degeneracy. Verification against real GRBs with secure spectroscopic redshifts tests whether the simulation-trained network generalizes to real data.

\begin{acknowledgments}
We thank Prof. Shao-Lin Xiong for helpful discussions and valuable suggestions on the manuscript.
\end{acknowledgments}

\bibliography{ApJ-refs_lobi2}

\end{document}